\documentclass[prl,twocolumn,showpacs,amssymb,amsmath,superscriptaddress]{revtex4}

\usepackage[dvips]{graphicx}
\usepackage{dcolumn}
\usepackage{bm}

\begin{document}

\title{Constraining a possible time-variation of the gravitational constant
through ``gravitochemical heating'' of neutron stars}

\author{Paula Jofr\'e}
\author{Andreas Reisenegger}%
\affiliation{Departamento de Astronom\'\i a y Astrof\'\i sica,
Pontificia Universidad Cat\'olica de Chile, Casilla 306, Santiago 22, Chile}%

\author{Rodrigo Fern\'andez}
\affiliation{Department of Astronomy \& Astrophysics, University of Toronto, Toronto, ON M5S 3H8, Canada.}%
\date{\today}

\begin{abstract}
A hypothetical time-variation of the gravitational constant $G$
would cause neutron star matter to depart from beta equilibrium, due
to the changing hydrostatic equilibrium. This forces non-equilibrium
beta processes to occur, which release energy that is invested
partly in neutrino emission and partly in heating the stellar
interior. Eventually, the star arrives at a stationary state in
which the temperature remains nearly constant, as the forcing
through the change of $G$ is balanced by the ongoing reactions.
Comparing the surface temperature of the nearest millisecond pulsar,
PSR~J0437-4715, inferred from ultraviolet observations, with our
predicted stationary temperature, we estimate two upper limits for
this variation: (1) $|\dot G/G| < 2 \times 10^{-10}$ yr$^{-1}$, if
we allow direct Urca reactions operating in the neutron star core,
and (2) $|\dot G/G| < 4 \times 10^{-12}$ yr$^{-1}$, considering only
modified Urca reactions. Both results are competitive with those
obtained by other methods, with (2) being among the most
restrictive.
\end{abstract}


\maketitle


Since \citet{dirac37} suggested that the gravitational force may be weakening with the expansion
of the universe, the question about whether or not the fundamental constants of nature vary in time
has been of interest. General relativity assumes a strictly constant gravitational coupling
parameter $G$, but alternatives such as the Brans-Dicke scalar-tensor theory
\citep{brans-dicke61} predict a variable $G$.

There have been many experiments attempting to test the constancy of
$G$, obtaining upper limits for its variation, most usefully
expressed as $| \dot G / G |$.  The experiments can be separated
into three classes.
In the first class are those that measure the variation of $G$ from
the early Universe to the present time, over timescales $\gtrsim
10^{10}$~yr, such as Big-Bang nucleosynthesis \cite{copi04} and
anisotropies in the cosmic microwave background radiation
\cite{nagata04}, yielding upper limits on the long-term averaged
variation, $| \dot G / G | \lesssim 10^{-13}$~yr$^{-1}$.
The second class contains measurements on long timescales,
$10^{9-10}$~yr, but without reaching into the very early Universe.
In this category, one finds constraints based on paleontology
\cite{eichendorf77} and on stellar astrophysics, the latter
including helioseismology \cite{guenther98}, binary pulsar masses
\cite{thorsett96}, and globular clusters \cite{innocenti95}.
Inferred constraints are
in the range $| \dot G / G | \lesssim
10^{-(11-12)}$~yr$^{-1}$.
Experiments in the third class measure the change of $G$ directly
over short, ``human'' timescales, $\sim 10$~yr. Orbits of planets
\cite{williams96}, rotation of isolated pulsars \cite{goldman90},
orbits of binary pulsars \cite{kaspi94}, and oscillations of white
dwarfs \cite{benvenuto04} are some examples of this class, with
inferred upper limits $| \dot G / G | \lesssim
10^{-(10-12)}$~yr$^{-1}$. Even though results from the first
category are nominally the most restrictive on a long-term variation
of $G$, they depend on the assumed form of the variation of $G$ near
the Big Bang. Thus, it is still useful to consider measurements of
the second and third categories, which could detect variations of
$G$ in more recent times.

In this paper, we present a new method for constraining a possible
time-variation of $G$, which we call \emph{gravitochemical heating}.
A change in $G$ induces a variation in the internal composition of a
neutron star, causing dissipation and internal heating. The
comparison of the predicted surface temperatures with the only
available observation~\cite{kargaltsev04} sets constraints on the
variation of $G$. The method is based on the results of Fern\'andez
\& Reisenegger~\cite{fernandez05} (see also
Refs.~\cite{reisenegger95}, \cite{reisenegger97}, and
\cite{reisenegger06}), who showed that internal heating arises as a
consequence of the compression of the star due to the decrease of
its rotation rate and centrifugal force (\emph{rotochemical
heating}). In the present work, compression (or expansion) of the
star is caused by an increase (or decrease) in the gravitational
force as $G$ changes, with the governing equations having the same
mathematical form. Since the chemical relaxation timescale of a
neutron star is $\sim 10^8$~yr, our experiment is closest to those
in the second class discussed above. If $G$ changes over
cosmological timescales (as in the proposals of Dirac and
Brans-Dicke) $\dot G$ can be approximated as a constant, which we
will do in most of what follows.


\emph{Theory.--} The simplest neutron-star models contain only
neutrons ($n$), protons ($p$), and low-mass leptons ($l$; including
electrons [$e$] and muons [$\mu$]), which can convert into each
other by beta reactions: direct Urca,
\begin{eqnarray}
\label{urca}
n & \longrightarrow & p + l + \overline{\nu}, \nonumber \\
p + l & \longrightarrow & n + \nu,
\end{eqnarray}
possibly forbidden by momentum conservation, and modified Urca,
\begin{eqnarray}
\label{murca}
n + N & \longrightarrow & p + N + l + \overline{\nu}, \nonumber \\
p + l + N  &  \longrightarrow & n + N + \nu,
\end{eqnarray}
where an aditional nucleon $N$ must be introduced in order to
conserve momentum. Neutrinos ($\nu$) and antineutrinos
($\overline{\nu}$) produced in these reactions escape,
taking energy with them and therefore cooling the
star~\cite{yakovlev04}. When the latter
is in beta equilibrium
(with the chemical potentials satisfying $\mu_n = \mu_p + \mu_l$,
for both $l=e,\mu$), the rate of direct and inverse reactions is the
same, i.e., the composition does not change.

A change in $G$ produces a change in the density of the stellar
matter. Since the chemical potentials depend on density, the system
departs from the beta equilibrium state. This departure can be
quantified by the chemical imbalances,
\begin{equation}
\eta_{npl} = \mu_{n} - \mu_{p} - \mu_l.
\end{equation}
It increases the reaction rates so as to approach a new equilibrium
configuration. If $G$ changes continuously, the star is always out
of equilibrium, storing an excess of energy that is dissipated as
internal heating and enhanced neutrino emission.

The formalism for calculating the evolution of the temperature and
chemical imbalances is described in \S~2 of Ref.~\cite{fernandez05}.
Here we outline the fundamental equations and the modifications
required in order to treat the gravitochemical case. The evolution
of the internal temperature as measured by a distant observer, $T$
(we avoid the usual subscript ``$\infty$'' in order to keep the
equations simple), taken to be uniform inside the star, is given by
the thermal balance equation,
\begin{equation}
\label{dotT} \dot T=\frac{1}{C} \left(L_H-L_{\nu}-L_{\gamma}
\right),
\end{equation}
where $C$ is the total heat capacity of the star, $L_H$ is the total
power released by the heating mechanism (which depends on the
chemical imbalances and the temperature), $L_{\nu}$ the total power
emitted as neutrinos, and $L_{\gamma}$ the power released as thermal
photons. The evolution of the chemical imbalances is given by
\begin{eqnarray}
\label{doteta} \dot{\eta}_{npl} & = & - \left[A_{D,l}(\eta_{npe},T)
+
A_{M,l}(\eta_{npe},T)\right]\nonumber \\
 & & - \left[B_{D,l}(\eta_{np\mu},T)+ B_{M,l}(\eta_{np\mu},T)\right]
+ C_{npl} \dot G,
\end{eqnarray}
where the subscripts $D$ and $M$ refer to direct and modified Urca
reactions, respectively. The functions $A$ and $B$ quantify the
effect of reactions towards restoring chemical equilibrium, and thus
have the same sign of $\eta_{npl}$ \cite{fernandez05}. The constant
$C_{npl}$ quantifies the departure from equilibrium due to $\dot G$,
being positive and depending on the stellar model and equation of
state. It can be further decomposed as $C_{npl} =
(Z_{npl}-Z_{np})I_{G,l}+Z_{np}I_{G,p}$, where $I_{G,i} \equiv
(\partial N_i^{eq}/\partial G)_A$ is the change of the equilibrium
number of each particle species, $N_i^{eq}$, due to the variation of
$G$, at constant total baryon number $A$, while $Z_{i}$ are
coefficients depending only on the stellar model (see
Ref.~\cite{reisenegger06} for a rigorous calculation of the latter).


As in the rotochemical heating case, the main consequence of
gravitochemical heating is that the star arrives at a stationary
state, where the rate at which $\dot G$ modifies the equilibrium
concentrations is the same as the rate at which reactions drive the
system toward the new equilibrium configuration, with heating and
cooling balancing each other \citep{reisenegger95}. The properties
of this stationary state can be obtained by the simultaneous
solution of equations (\ref{dotT}) and (\ref{doteta}) with $\dot T =
\dot \eta_{npl} = 0$. The existence of the stationary state makes it
unnecessary to model the full evolution of the temperature and
chemical imbalances of the star in order to calculate the final
temperature, since the stationary state is independent of the
initial conditions (see Ref.~\cite{fernandez05} for a detailed
analysis of the rotochemical heating case). For a given value of
$|\dot G|$, it is thus possible to calculate the temperature of an
old pulsar that has reached the stationary state, without knowing
its exact age.

When only modified Urca reactions operate and nucleon Cooper pairing
effects are negligible, it is possible to solve analytically for the
stationary values of the photon luminosity $L_{\gamma}^{st}$ and
chemical imbalances $\eta_{npl}^{st}$, as a function of stellar
model and current value of $|\dot G/G|$. The reason for this is that
the longer equilibration timescale given by the slower modified Urca
reactions yields stationary chemical imbalances satisfying
$\eta_{npl}\gg kT$. The term $L_H-L_{\nu}$ in the thermal balance
equation can be written asymptotically as $K_{Le}\eta_{npe}^8 +
K_{L\mu} \eta_{np\mu}^8$, where $K_{L,l}$ are positive constants
that depend only on stellar mass and equation of
state~\cite{fernandez05}. Assuming a stellar model with mass $1.41
M_\odot$, calculated with the equation of state A18+$\delta
\upsilon$+UIX* \cite{apr98}, the photon luminosity in the stationary
state comes out to be
\begin{equation}
L_{\gamma}^{st} \simeq 5 \times 10^{28} \left(|\dot G/G|\over
10^{-12} \textrm{\,\ yr}^{-1}\right)^{8/7} \textrm{erg s}^{-1}.
\end{equation}
Assuming isotropic blackbody radiation, i.e., $L_{\gamma} = 4 \pi
\sigma R^2 T_s^4$, with $R\simeq 14~{\rm km}$ the radius of the star
as measured by a distant observer, and $\sigma$ the Stefan-Boltzmann
constant, the surface temperature of the star in the stationary
state is
\begin{equation}
T_{s}^{st} \simeq 7.7 \times 10^4 \left(\frac{| \dot G/G
|}{10^{-12}\textrm{\,\ yr}^{-1}}\right)^{2/7} \textrm{K}.
\end{equation}
The timescale for the system to reach the stationary state, obtained
by the same procedure as in \S~4.4 of Ref.~\cite{fernandez05}, is
\begin{equation} \label{tau_eq}
\tau_{st} \simeq 2 \times 10^8 \left(\frac{| \dot G/G
|}{10^{-12}\textrm{\,\ yr}^{-1}}\right)^{-6/7} \textrm{yr}.
\end{equation}


\emph{Results.--} In order to apply this formalism and constrain the
value of $|\dot G/G|$, we need a neutron star that (1) has a
measured surface temperature (or at least a good enough upper limit
on the latter), and (2) is confidently known to be older than the
timescale to reach the stationary state for the intended upper limit
on $|\dot G/G|$. So far, the only object satisfying both conditions
is the millisecond pulsar closest to the Solar System,
PSR~J0437-4715 (hereafter ``J0437''), whose surface temperature was
inferred from an ultraviolet observation by Kargaltsev, Pavlov, and
Romani~\cite{kargaltsev04}. Its spin-down age, $\tau_\mathrm{sd}
\simeq 5 \times 10^{9}$~yr (e.g., \cite{van_straten01}), and the
cooling age of its white dwarf companion, $\tau_\mathrm{WD}\simeq
2.5-5.3 \times 10^{9}$~yr~\cite{hansen98}, are much longer than the
time required to reach the steady state, as long as $| \dot G / G |
\gtrsim 10^{-13}$ yr$^{-1}$.

\begin{figure}
\includegraphics[scale=0.525]{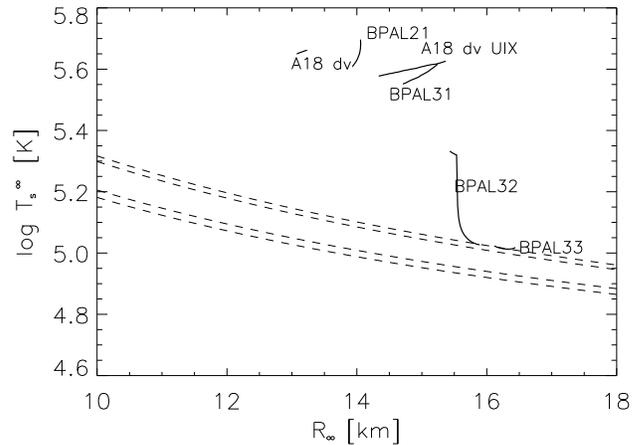}
\caption{Comparison of the gravitochemical heating predictions to
observations of PSR~J0437-4715. The solid lines are the predicted
stationary surface temperatures as functions of stellar radius, for
different equations of state (A18 from \cite{apr98} and BPAL from
\cite{pal88}), constrained to the observationally allowed mass range
for this pulsar~\cite{van_straten01}. Dashed lines correspond to the
68\% and 90\% confidence contours of the blackbody fit of Kargaltsev
et al.~\cite{kargaltsev04} for the ultraviolet emission from this
object. The value of $|\dot{G}/G| = 2 \times 10^{-10}$~yr$^{-1}$ is
chosen so that all stationary temperature curves lie above the
observational constraints. (BPAL32 and BPAL33 allow direct Urca
reactions in the observed mass range of J0437.) \label{J0437_1}}
\end{figure}

Consequently, we consider stellar models constructed from different
equations of state, with masses satisfying the constraint obtained
for J0437 by van Straten et al.~\cite{van_straten01},
$M_\mathrm{psr} = 1.58 \pm 0.18 M_{\odot}$, and calculate the
stationary temperature for each, ignoring potential Cooper pairing
effects~\cite{yakovlev04}. In Figure~\ref{J0437_1}, we compare the
observational constraints on the temperature of this pulsar
~\cite{kargaltsev04} together with the theoretical prediction for
different equations of state (for further details see Table 1 of
Ref.~\cite{fernandez05}), assuming $|\dot{G}/G| = 2\times
10^{-10}$~yr$^{-1}$. When the stellar mass becomes large enough for
the central pressure to cross the threshold for direct Urca
reactions, $T_s$ drops abruptly, due to the faster relaxation
towards chemical equilibrium. This occurs in two steps, as electron
and muon direct Urca processes have different threshold densities
(see, e.g., Ref.~\cite{fernandez05}). The chosen value of $\dot G$
is such that the stationary temperatures of all stellar models lie
(just) above the 90 \% confidence contour, as seen in
Figure~\ref{J0437_1}, and therefore represents a rather safe,
general upper limit.

\begin{figure}
\includegraphics[scale=0.525]{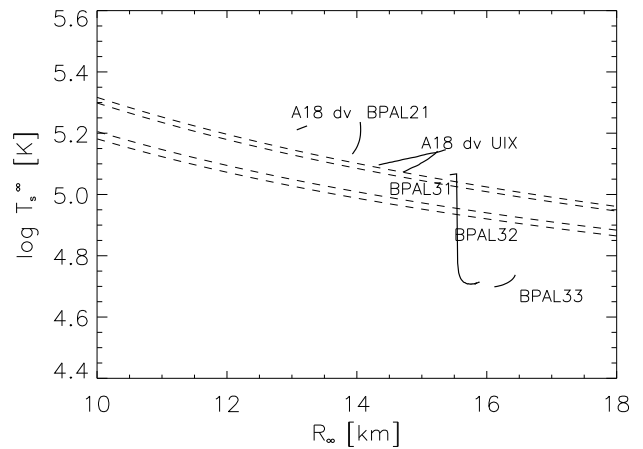}
\caption{Same as Figure~\ref{J0437_1}, but now the value of
$|\dot{G}/G| = 4 \times 10^{-12}$ yr$^{-1}$ is chosen so that only
the stationary temperature curves with modified Urca reactions are
above the observational constraints. \label{J0437_2}}
\end{figure}

Nevertheless, conventional neutron star cooling models reproduce
observed temperatures better when only modified Urca reactions are
considered~\cite{yakovlev04}. Restricting our sample to the
equations of state which allow only modified Urca reactions in the
mass range considered here, namely A18 + $\delta v$, A18 + $\delta
v$ + UIX, BPAL21, and BPAL31, we obtain a more restrictive upper
limit, $| \dot G / G | < 4 \times 10^{-12}$ yr$^{-1}$, illustrated
in Figure~\ref{J0437_2}. This value is among the most restrictive
results from methods probing a similar timescale.


So far, we have assumed a steady, monotonic time-variation of $G$.
Can our method constrain short-timescale variations? We consider an
oscillatory variation, $G(t) = G_0 [1+A\cos(\omega t)]$, in a star
with only modified Urca reactions and one leptonic species
(electrons), in the interesting limit $\eta_{npe}\gg kT$. The
equation for the chemical imbalance evolution is
\begin{equation}
\label{eta_osc} \dot \eta_{npe} = - K_A \eta_{npe}^7 + C_{npe} \dot
G.
\end{equation}
In the regime of low forcing frequency (when the period is longer
than the timescale to reach the steady state), this evolution is
equivalent to the situation already considered. In the opposite
limit, when the oscillation is faster than the time required to
reach the steady state, we have $\eta_\mathrm{npe} \simeq C_{npe} A
G_0 \cos (\omega t)$. As mentioned above, the photon luminosity
satisfies $L_\gamma \simeq K_{Le} \eta_\mathrm{npe}^8$. Assuming a
detection limit similar to the luminosity observed in J0437,
$L_\gamma\sim 10^{29}~\mathrm {erg~s^{-1}}$, the heating can be
detected only if the oscillation timescale satisfies
\begin{equation}
t_\mathrm{osc} \equiv {1\over\omega}\sim
\frac{(L_\gamma/K_{Le})^{1/8}}{C_\mathrm{npe}|\dot{G}|}\gtrsim
3\times 10^8 \left(\frac{|\dot{G}/G|}{10^{-12}\textrm{
yr}^{-1}}\right)^{-1}\textrm{ yr},
\end{equation}
too long to be of interest.

\emph{Conclusions.--} We have shown that a long-term, monotonic
variation of $G$ alters the chemical equilibrium state of a neutron
star, leading to internal heating. This effect is an exact analog of
that produced by spin-down compression~\cite{fernandez05}. As in
that case, the star reaches a stationary state in which the
temperature remains nearly constant. We have calculated this
temperature as a function of $|\dot{G}/G|$ and stellar mass using
realistic equations of state, and compared our results to the
temperature inferred from ultraviolet observations of the
millisecond pulsar J0437-4715 by Kargaltsev et
al.~\cite{kargaltsev04}. In the most general case, when direct Urca
reactions are allowed to operate, we obtain an upper limit $|\dot G
/ G| < 2 \times 10^{-10}$ yr$^{-1}$. Restricting the sample of
equations of state to those that allow only modified Urca reactions,
we obtain a much more restrictive upper limit, $|\dot G / G| < 4
\times 10^{-12}$ yr$^{-1}$, competitive with constraints obtained
from other methods probing similar timescales. However, since the
composition of matter above nuclear densities is uncertain and
millisecond pulsars are generally expected to be more massive than
classical pulsars, we cannot rule out the result for the direct Urca
regime. Information about the prevalence of the direct Urca process
and the mass threshold above which it is present may in the future
be obtained from the study of the quiescent emission of soft X-ray
transients (e.~g., Ref.~\cite{yakovlev04}).

Further progress in our knowledge of neutron star matter will allow
this method to become more effective at constraining variations in
$G$. The method can also be improved with an increased sample of
objects with measured thermal emission. In this respect, an
interesting candidate is PSR~J0108-1431, plausibly the nearest
pulsar, with distance estimates ranging from 60~pc~\cite{tauris94}
to 184~pc~\cite{ne2001}. Due to its much slower rotation (period
$P=0.808~{\rm s}$~\cite{tauris94}, compared to $P=0.00576~{\rm s}$
for J0437~\cite{van_straten01}), the current time-derivative of its
centrifugal force ($\propto\dot P/P^3$) is 680 times smaller than
that of J0437. Thus, unless it was born rotating much faster
($P_0\lesssim 0.02~{\rm s}$), rotochemical heating should be
negligible for this pulsar~\cite{fernandez05,gonzalez}, and any
detectable surface temperature would be a clear signature of a
change in $G$.

\acknowledgments This work was supported by FONDECYT (Chile) through
its Regular Grant 1060644.


\end{document}